\documentclass[11pt]{amsart}
\usepackage{amscd,amssymb,amsmath,latexsym,enumerate}
\usepackage{bbm}
\usepackage[mathscr]{euscript}
\usepackage{epsfig}
\DeclareMathAlphabet{\mathpzc}{OT1}{pzc}{m}{it}

\newcommand{\Dd}{{\mathcal D}}

\newcommand{\Hh}{{\mathcal H}}

\newcommand{\RM}{{\mathbb R}}

\newcommand{\hp}{{\widehat{p}}}

\newcommand{\hx}{{\widehat{x}}}

\let\X\chi
\renewcommand{\chi}{\raisebox{2pt}{$\X$}}

\begin{document}

\title{Comment on ``Hamiltonian for the Zeros of the Riemann Zeta Function''}
\author{Jean V Bellissard}

\address{Georgia Institute of Technology\\
School of Mathematics\\
Atlanta GA, USA}
\email{jeanbel@math.gatech.edu}


\maketitle


\noindent In a recent paper published in PRL, C.~M.~Bender, D.~C.~Brody, and M.~P.~M\"uller \cite{BBM} propose a strategy to show that the non trivial zeroes of the Riemann $\zeta$-functions lie on the line $\Re{z}=1/2$. The strategy is based on the Hilbert-Polya conjecture that these zeroes can be seen as the eigenvalues of a selfadjoint operator. Inspired by previous seminal works, in particular the ones by Berry and Keating \cite{BK} on the one hand and the one by Connes \cite{Co99} (see also \cite{CM08}, especially Chapter 2, Section 3) on the other hand, they define an operator $H$ similar to the quantum analog of the generator of the dilation operator $A=(\hx\hp+\hp\hx)/2$. By {\em similar} it is meant that $H$ is obtained from $A$ by a change of basis that is not unitary. However, extra discrete symmetries of $H$ are used to show that these eigenvalues are real. The main idea can be summarized as follows. If $\Re{z}>1$,

\begin{equation}
\label{rh17.eq-psi}
\psi_z(x) = \sum_{n=1}^\infty \frac{1}{(n+x)^z}
   \hspace{1cm}\Rightarrow\hspace{1cm}
     \psi_z(x)-\psi_z(x-1)=-1/x^z\,,
      \;\; x> 0\,.
\end{equation}

\noindent The function $\psi_z$ is well defined on the interval $x\in (-1,+\infty)$, but it is singular for $x\leq -1$, where, depending on the value of $z$ it has either poles or cuts. Like the Riemann $\zeta$-function, $\psi_z$ admits an integral representation which allows to define it as a meromorphic function on $\Re{z}>0$ with a simple pole at $z=1$

\begin{equation}
\label{rh17.eq-ext}
\psi_z(x)= \frac{1}{(z-1)\Gamma(z)}
   \int_0^\infty
    \frac{t^{z-1}e^{-tx}}{4\sinh^2(t/2)}\,
     \left\{
      1-t-e^{-t}-xt(1-e^{-t})
     \right\}\;dt
\end{equation}

\noindent Since the map $z\to x^{-z}$ is analytic in $\Re{z}>0$ when $x>0$, it follows that eq.~(\ref{rh17.eq-psi}) still holds for $\Re{z}>0$ as meromorphic functions and $x>0$. Since the function $x\in (0,+\infty)\mapsto 1/x^z$ is an eigenfunction for $A$ with eigenvalue $-\imath (z-1/2)$, it follows that $\psi_z$ is a candidate to be an eigenfunction of the operator obtained from $A$ by the operator $\Delta$ defined by $\Delta f=f(x)-f(x-1)$ for $x>0$. If this operator is invertible then $H=\Delta A\Delta^{-1}$ gives $H\psi_z=\imath (z-1/2)\psi_z$. In order to get the zeroes of the $\zeta$-function it is sufficient to impose a Dirichlet boundary condition at $x=0$, because $\zeta(z)=\psi_z(0)$. Then $H$ is not selfadjoint, but the authors show that it admits symmetries that force its eigenvalues to be real leading to a potential proof of the Riemann Hypothesis. While the idea is appealing, a closer look at the paper raises several problems that are not addressed and that make this approach quite questionable.

\vspace{.3cm}

\noindent {\bf Problem 1:} The authors quote: {\em ``The choice of the boundary condition $\psi_z(0)=0$ as discussed below, is motivated by our requirement that $\hp$ should be symmetric.''}. At this point, we need to lift a first ambiguity: what is the Hilbert space on which these objects, wave functions and operators, are defined~? A first guess would be to take $L^2(\RM)$, on which both the position $\hx$ and the momentum $\hp=-\imath d/dx$ are well defined and selfadjoint. But the Dirichlet boundary condition at $x=0$, on the one hand, and the expression of $\psi_z$ on the other hand, seems to indicate that $\Hh=L^2(0,+\infty)$ is a more reasonable choice. The latter is also a good space on which the operator $A$ is well defined and self adjoint since it is the generator of the dilation operator $D(\lambda)f(x)=(1/\lambda)^{1/2} f(x/\lambda)$, which gives a strongly continuous unitary representation of the multiplicative group $\RM_+^\ast$ of positive real numbers on $\Hh$. However, the first problem arise: how is $\Delta$ defined~? For $\Delta f$ involves the translation $f\to f(\cdot -1)$ which is defined only for $x\geq 1$. The way the author propose to look at this problem is to represent this translation by $e^{\imath \hp}$ (actually $e^{-\imath \hp}$ with the conventions adopted for the definition of $\hp$), which is fine if $\hp$ can be proved to be selfadjoint. 

\vspace{.1cm}

\noindent At this point, a technical remark is in order here. Since $\hp$ is unbounded, it is necessary to define its domain $\Dd$. The standard choice is to choose the set of $f\in \Hh$ with distributional derivative in $\Hh$. Then the operator $\hp^\dag: f\to -\imath f'$ admits a closed graph in $\Hh\times \Hh$, since the graph norm is given by $\|f\|_H^2=\|f\|_{L^2}^2+\|f'\|_{L^2}^2$, defining a complete Hilbert space (Sobolev norm). It is an elementary exercise to prove that if $f\in \Dd$ then $|f(0)|\leq \|f\|_H$, showing that the Dirichelt boundary condition $f(0)=0$ defines a closed subspace $\Dd_0$ of the graph of $\hp^\dag$. Then the restriction of $\hp^\dag$ to $\Dd_0$ defines exactly the operator $\hp$ proposed in the article. That $\hp$ is symmetric is an exercise requiring an integration by part. But it is a classical result that its adjoint is $\hp^\ast=\hp^\dag$, which, by the same argument, is not symmetric.

\vspace{.1cm}

\noindent Actually, the spectrum of $\hp^\dag$ contains all points of the upper half plane as eigenvalues. This is because the functions $g_z(x)=e^{-zx}$ belong to $\Hh$ if and only if $\Re{z}>0$ and they satisfy $\hp^\dag g_z= -\imath g_z'= \imath z\, g_z$. However, the same argument shows that no point in the lower half plane belong to the spectrum. Actually the resolvent $(z-\hp^\dag)^{-1}$ can be computed explicitly if $\Im{z}<0$, by solving the differential first order equation $zf+\imath f'=h$ with $h\in \Hh$ and showing that all solutions are in $\Dd\subset \Hh$. Von Neumann studied the question of whether or not a symmetric operator $H$ admit selfadjoint extensions. He proved that the answer is yes if and only if the eigenspaces of the adjoint $H^\ast$ for the eigenvalues $\pm \imath$ have same dimension, called the {\em defect indices} $n_\pm$. If yes all selfadjoint extension are classified by the set of unitary operators between these two spaces \cite{Fa75}. But in the present case $\hp$ admits $n_+=1, n_-=0$, so that $\hp$ has no selfadjoint extension. Hence the argument proposed in the paper cannot be used.

\vspace{.1cm}

\noindent Another way to understand why it is so, is to consider the translation operator $S$ given by $Sf(x)=f(x-1)$. If defines on $L^2(\RM)$, $S$ is unitary. But if restricted to $\Hh=L^2(0,+\infty)$ it is not. It is only a partial isometry. Its adjoint $S^\ast$ satisfies $S^\ast S= I$ but $SS^\ast=I-P$ where $P$ is the projector onto $L^2(0,1)$. And $S^\ast$ admits every point inside the unit disc as an eigenvalue. The corresponding eigenvectors are, not surprisingly, related to the function $g_z$ above.

\vspace{.3cm}

\noindent {\bf Problem 2:} Using eq.~(\ref{rh17.eq-ext}), it becomes possible to compute the norm $\|\psi_z\|_{L^2}$. This expression can be computed as a double integral with singularities at $0$ and at $+\infty$. The exponential decay at infinity gives convergence. But the convergence near the origin holds only for $\Re{z}>3/2$. This can be seen already by using eq.~(\ref{rh17.eq-psi}). The meromorphic extension to the domains $\Re{z}>0$ does not actually help. Hence the {\bf wave function $\psi_z$ is not an eigenvector for $\Re{z}=1/2$}.

\vspace{.1cm}

\noindent One way to cure this problem could be to replace $\Hh$ by a weighted space $\Hh_\alpha$ with $\alpha\in \RM$ and norm

$$\|f\|_\alpha^2=
   \int_0^\infty |f(x)|^2\, x^\alpha dx\,.
$$

\noindent This is because the convergence at the origin holds if $\Re{z}>(3+\alpha)/2$. So taking $\alpha$ sufficiently negative insures that $\psi_z\in\Hh_\alpha$ at least if $\psi_z(0)=0$. Such a Hilbert space still support a unitary representation of the dilation group if one sets 

$$D(\lambda)f(x)= 
   \frac{1}{\lambda^{(1+\alpha)/2}}\,
    f\left(\frac{x}{\lambda}\right)\,.
$$

\noindent However the generator $A_\alpha$, which is selfadjoint by construction, becomes 

$$A_\alpha f(x) =
   \imath x\frac{d}{dx}+\imath \frac{\alpha +1}{2}\,.
$$

\noindent The constant term on the {\em r.h.s.} changes the eigenvalue problem and does not give the line $\Re{z}=1/2$ anymore. 

\vspace{.1cm}

\noindent {\bf Conclusion: } As attractive this idea looks, it does not hold when checking the analysis part of the problem.

\end{document}